\input harvmac
\input epsf
\parindent=.31truein
\hoffset=0truein
\voffset=-.1truein
\hsbody=\hsize \hstitle=\hsize 
\def\ra{\rightarrow}
\def\CM{{\cal M}}
\font\cmss=cmss10 \font\cmsss=cmss10 at 10truept
\def\IR{\relax{\rm I\kern-.18em R}}
\font\cmss=cmss10 \font\cmsss=cmss10 at 10truept
\def\IZ{\relax\ifmmode\mathchoice
{\hbox{\cmss Z\kern-.4em Z}}{\hbox{\cmss Z\kern-.4em Z}}
{\lower.9pt\hbox{\cmsss Z\kern-.36em Z}}
{\lower1.2pt\hbox{\cmsss Z\kern-.36em Z}}\else{\cmss Z\kern-.4em Z}\fi}
$\,$
\overfullrule=0pt
\Title{\vbox{\baselineskip12pt\hbox{
}\hbox{}}}
{\vbox{\centerline{Renormalization Ambiguities in Chern-Simons
Theory}}}

\centerline{\bf M. Asorey, F. Falceto,
J. L. L\'opez and G. Luz\'on {$^1$}}
\bigskip\centerline{{
Departamento de F\'{\i}sica Te\'orica.
 Facultad de Ciencias}} \centerline{Universidad de Zaragoza.
50009 Zaragoza Spain}

\baselineskip=16pt plus 2pt minus 1pt
 \bigskip
\bigskip\medskip
\centerline {\bf Abstract}
We introduce a new family of gauge invariant regularizations
of Chern-Simons theories which generate  one-loop renormalizations
of the coupling  constant of the form $k\rightarrow k+2 s c_v$ where
 $s$ can take any arbitrary  integer value. In the particular
case $s=0$ we get an explicit example of a gauge invariant regularization
which does not generate radiative corrections to the bare coupling
constant.  This ambiguity in  the radiative corrections
to $k$ is reminiscent of the Coste-L\"uscher  results
for the parity anomaly in (2+1)  fermionic  effective actions.
\overfullrule=0pt
\hyphenation{systems}  \bigskip\vfill \noindent\baselineskip=12pt
plus 2pt minus 1pt
\vfill
\noindent
PACS numbers, 11.10 Kk, 11.15.-q, 11.25.Hf\hfill\break
\Date{ }
\parindent=20pt
\baselineskip=16pt plus 2pt minus 1pt
Although the Chern-Simons theory is exactly solvable in the
canonical formalism, there are still many interesting open questions
related to the behavior of the theory in  arbitrary
three-dimensional manifolds which can only be answered in the
covariant formalism. In particular, Witten's  conjectures
on the connection of expectation values of gauge
invariant observables with topological invariants in arbitrary
three-dimensional manifolds \ref\witten{E. Witten,
Commun. Math. Phys. 121 (1989) 351
} can only be formulated in a
covariant formalism. Some of the conjectures have been proven
by different methods such as surgery of manifolds, operator
formalism or Wess-Zumino-Witten model approaches  with a few
additional assumptions.
 However, from a field theoretical point of
view is still a challenge  to prove those conjectures by  functional
integral methods \ref\axe{ S. Axelrod, I.M. Singer,  Proceedings of the XXth
International Conference on Differential Geometrical Methods in Theoretical
Physics, S. Catto and A. Rocha eds., World Sci. (1991)}\nref\freed{D. Freed, R.
Gompf, Commun. Math. Phys. { 141} (1991) 79}--\ref\natan{D. Bar-Natan, E.
Witten,  Commun. Math. Phys. { 141} (1991) 423}. One of the major problems of
this approach
is the presence of ultraviolet divergences which require the
introduction of a regularization. All gauge invariant  regularization
methods considered so far  yield  a finite radiative correction
to the coupling constant $k$ of the form $k\rightarrow k+ c_v$.
This fact has been interpreted as an explanation of the
appearance of  the expression $ k+ c_v$ in many
formulae in representation theory of the corresponding
Kac-Moody algebra. However, we will see in this paper that the
claims about the universality of  this correction  do not hold in more general
regularization schemes.

Most of the gauge invariant regularizations  introduce scalar selfinteracting
terms  of Yang-Mills type for  gluons while keep the scalar
character of  ghost-gauge field interactions
\ref\sem{W. Chen,
G. W. Semenoff, Y.-S. Wu, Mod. Phys. Lett. A5  (1990)
1833}\nref\LAG{L. Alvarez-Gaum\'e, J.M.F. Labastida, A.V.
Ramallo,  Nucl. Phys. {
B334} (1990) 103}\nref\pl{M. Asorey, F. Falceto,  Phys. Lett. B 241 (1990) 31
}--\ref\CM{C.P.  Martin,
Phys. Lett. { B 241} (1990) 513}.
In the regularizations of this type the effective coupling constant $k$  gets a
finite radiative correction of the form $ k+ c_v$ at one loop and nothing else
at two loops \ref\car
{G. Giavarini,  C.P.  Martin, F. Ruiz Ruiz, Nucl.Phys.
{ B381}(1992) 222}. This phenomenon
indicates  that the effective values of $k$ are very
strictly constrained and suggests that this behavior might be
universal.

{}From a pure perturbative viewpoint any
restriction on the renormalization of a coupling constant has to be
understood in terms of symmetry arguments.  In this case, however,
there is not such a symmetry to explain the  behavior of the
quantum correction to $k$.
The invariance of the bare action under large gauge transformations
implies that $k$ has to
be an integer, but this fact does not imposes any restriction on the value
of the renormalized coupling constant
(see Ref. \ref\new{ M. Asorey, F. Falceto,
J. L. L\'opez, G. Luz\'on, DFTUZ-93.10 preprint} for a general discussion
of the problem in Chern-Simons theory).

In this note we address the question of whether the above behavior is universal
or it is regularization dependent. In particular, we find out the existence of
ambiguities in the
renormalization of the Chern-Simons coupling constant which depend on the
regularization scheme. We shall
introduce a new family of continuum regularizations of the theory
by adding to the Chern-Simons  action new  higher covariant derivative
pseudoscalar terms  and
some Pauli-Villars regulators with scalar and pseudoscalar couplings. The
regularizations are inspired by the Coste-L\"uscher method of analyzing
the  ambiguities which appear in the effective action of
fermionic determinants \ref\lc{A. Coste, M.  L\"uscher, Nucl.Phys.
 { B323} (1989) 631}. Regularizations with  similar
characteristics can be obtained  by  geometric regularization, but
for the sake of simplicity we shall not consider them here
(see \new\ for details).

The Chern-Simons action
\eqn\act{S_0(A) ={ik\over4\pi}\int_{M}Tr(A\wedge
 dA +{2\over3}A\wedge A\wedge A) }
is invariant under diffeomorphim transformations
of the base manifold $M$ and global gauge transformations, provided
the coupling constant $k$ is an integer $k\in \IZ$. In the
functional integral quantization approach there are several possible
sources of symmetry breaking. First,  to define the functional
integral it is necessary to introduce a functional volume element
$[\delta A]$ which depends on the Riemannian structure of $M$. Another
possible source of symmetry breaking is the  gauge
fixing condition. A simple BRST analysis shows that both symmetries,
 gauge and
diffeomorphism invariance, can be preserved at the quantum level,
at the price of  generating  a framing  anomaly \witten\freed.
{}From a less formal point of view there is however another source of
symmetry breaking. The existence of ultraviolet divergences in the
covariant formalism makes necessary the introduction of
a regularization, and depending on the type of regularization some
of the classical symmetries may be broken.

 Since both symmetries cannot be simultaneously
preserved by an ultraviolet regularization we
will  try to preserve only one of them, namely, gauge invariance.

Most of the regularization methods based on  higher covariant
derivatives introduce  Yang-Mills like terms into the regularized
action which modify the pseudoscalar character of the original
Chern-Simons action. Here we shall consider a regularization
by higher covariant derivatives which preserves this peculiarity of
the Chern-Simons action.

Let
$\ast$ denote the Hodge operator associated to the
oriented Riemannian structure of M, and $d^{\ }_A$ and
$\Delta^{\ }_A=d^{\ast }_A d^{\ }_A + d^{\ }_A d^{\ast }_A$  the
covariant exterior differential and  the covariant laplacian operator,
respectively. The novelty of the present regularization is that the
regularized  action
\eqn\rega{	S_\Lambda(A) =S_0(A)-{ik\over8\pi\Lambda^2}\left(\ast
F(A), (1+{\Delta_A\over\Lambda^2})^n \ast
d_A(1+{\Delta_A\over\Lambda^2})^n \ast F(A)\right)}
 contains only pseudoscalar couplings like the original Chern-Simons action.
The bracket $(,)$ of the regulating term in \rega\
denotes the inner product defined by
\eqn\prod{(\alpha,\beta)=-2 \int_M \tr\  \alpha\wedge\ast\beta}
in the
space of forms $\alpha\  \beta$ taking values in the Lie
algebra of the (simple and compact) structure group $G$. The new
pseudoscalar term  was already introduced in Ref.
\pl\ for a different purpose. It
 is well known however that the method of higher covariant derivatives
does not
get rid of  one loop divergences which have to be eliminated by an
additional Pauli-Villars regularization.

The final expression  of the regularized functional integral in the
Landau gauge $d^\ast A=0$
\eqn\part{Z_\Lambda=\int [\delta A]\, [\delta
\phi]\,[\delta \bar {\psi}]\, [\delta \psi]\,[\delta
\bar c]\, [\delta c] \, \delta(d^\ast A) \exp\{-S_{reg}(A,\bar{c},
 {c},\phi, \bar {\psi}, {\psi})\}, }
is given in terms of the effective action
\eqn\rega{\eqalign{S_{reg}(A,\bar{c},
 {c},\phi, \bar {\psi}, {\psi})= S_\Lambda(A)+ &(\bar{c}, d^\ast_{\
} d^{\ }_A c)+{1\over 2} \sum_{i=1}^{M_b} ({\phi_i}, [I+
{\lambda_i\over \Lambda^2}\Delta^{\ }_A]^{m_i} \phi_i)\cr
&+
\sum_{j=1}^{M_f} (\bar{\psi_j}, [I+ {\mu_j\over \Lambda^2}\Delta^{\
}_A]^{n_j} \psi_j).\cr}}
where besides the Faddeev-Popov ghosts  $\bar{c}, {c}$ there appear
 bosonic
 Pauli-Villars ghosts   $\phi_i$ and grassmanian  fermionic ghosts
 $\psi_j$ and $\bar{\psi_j}$. The interactions of the
fields $\phi_i$, $\psi_j$ and
$\bar{\psi_j}$ with the gauge fields differ from gluon
selfinteractions, and therefore
the cancellation of the ultraviolet divergences does not proceed as
in  conventional Pauli-Villars method. In particular, the finite
radiative corrections may depend  on the prescription used to
calculate each diagram. If the manifold $M$ is a
three-dimensional torus, a very natural unambiguous prescription is
to introduce an auxiliary momentum pre-cutoff $|p|<\Omega$ for all
propagating modes. The only problem with this prescription is that
it breaks gauge invariance and it is therefore necessary to impose some
subsidiary conditions to ensure that it is restored  when the
pre-cutoff is removed ($\Omega\ra\infty$)\ref\sing{M. Asorey, F. Falceto,
 Int. J. Mod. Phys.{\bf 7}(1992) 235
}.

The only divergent contributions arise in the  two-point function
$\Gamma^{ab}_{\mu\nu}(q)$ and can be cancelled by an appropriate
choice of the Pauli-Villars regulators. We will restrict ourselves
to the case of a three-dimensional symmetric flat torus $M=T^3$ with
volume $L^3$. The leading contribution in the large $L$ limit is
$L$ independent and can be estimated by replacing the infinite
sums of propagating momentum modes by conventional integrals.

The radiative corrections to the two-point function generated by
one loop of gluons (diagrams (1) and  (2) of Fig. 1) are given by
\eqn\gluons{\eqalign { ^{^{(1)}}\Gamma_{\mu \nu}^{ab}(q)  = &
{c_v\over 6\pi^2}\, (16n^2+24n+9)\,\Omega\,\delta^{ab}\,\delta_{\mu\nu}
\cr & +
{2 c_v\over 3\pi^2}\,\Lambda\, I(n)\,\delta^{ab}\,\delta_{\mu\nu}  \cr & -
{c_v\over 32\pi^2}\, \left[(4n+3)^2+{\pi^2 \over 2}\right] \, \delta^{ab}\left (\vert q \vert
\delta_{\mu\nu} +
{q_{\mu}q_{\nu} \over \vert q \vert} \right )  \cr & +
\CO(\Omega^{-1},L^{-1},\Lambda^{-1}), \cr}
}
and
\eqn\gluonss{\eqalign{ ^{^{(2)}}\Gamma_{\mu \nu}^{ab}(q) = & -
{c_v\over 3\pi^2}(8n^2+14n+5)\,\Omega\,\delta^{ab}\,\delta_{\mu\nu}
\cr & - {2c_v\over 3\pi^2}\,\Lambda\, I(n)\,
\delta^{ab}\,\delta_{\mu\nu}  \cr &+ \CO(\Omega^{-1},L^{-1},\Lambda^{-1}), \cr
}
}
with
$$
I(n)=\int_{0}^{\infty}dp \, {\lbrack 5+2(5+9n)p^2 \rbrack\, p^2\,
(1+p^2)^{2n-2} \over 1+p^2(1+p^2)^{2n}} .
$$
Part of the
finite contribution which depends on $|q|$ comes from the existence
of the pre-cutoff $\Omega $ in every  propagator $|p|<\Omega$,
$|p+q|<\Omega$ of diagram (1) of Fig.1
\vskip.2cm
{\hskip-.8cm \epsfxsize=14cm \epsfbox{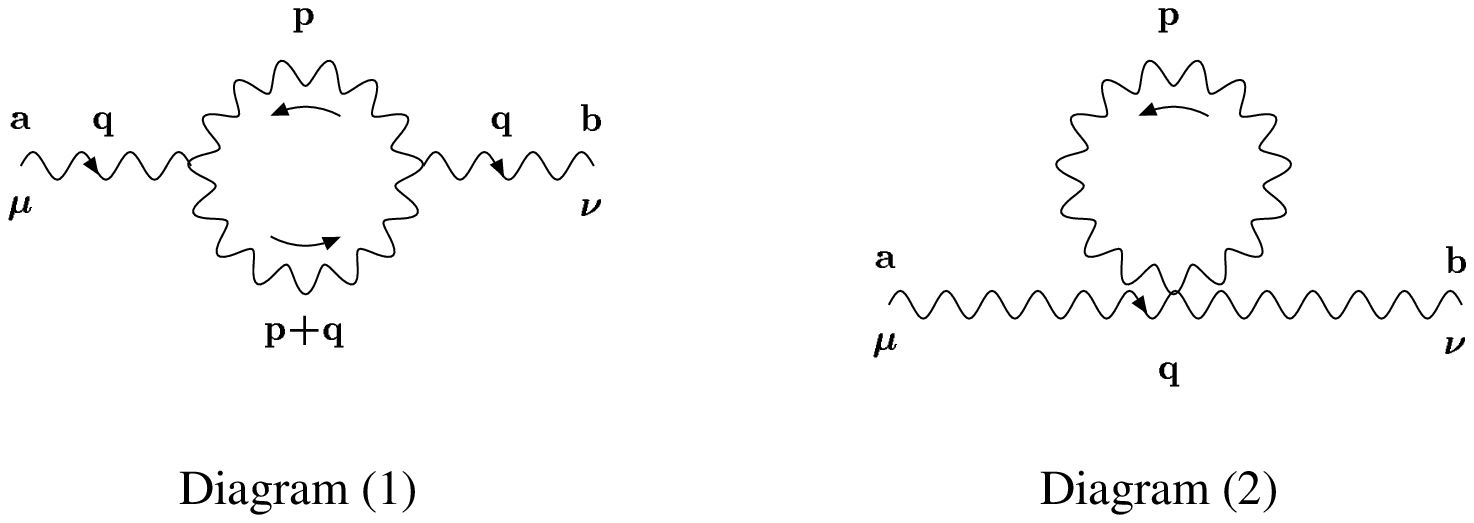}}
\vskip-.5cm
\vbox{\noindent {\bf Figure 1.}\it \baselineskip 10pt 
One loop Feynman diagrams contributing to the 2-point function
involving gluon loops.}
\baselineskip=16pt plus 2pt minus 1pt
%
$$\eqalign{\lim_{\Omega\ra\infty}&{c_v\over 8\pi^3}(4n+3)^2 \,\delta^{ab}\,
\sum_{|p|<\Omega\atop |p+q|>\Omega}{p^{8n}p_\mu p_\nu\over
[1+p^2(1+p^2)^{2n}]^2}\cr
&={c_v\over 32\pi^2}\,  (4n+3)^2
  \, \delta^{ab}\left (\vert q \vert
\delta_{\mu\nu}+
{q_{\mu}q_{\nu} \over \vert q \vert} \right ).  \cr}
$$
The contributions to the same proper function generated by the
Faddeev-Popov ghosts are given by
\eqn\fp{\eqalign {^{^{(FP)}}\Gamma_{\mu \nu}^{ab}(q)=& -
{c_v\over 6\pi^2}\,\Omega\, \delta^{ab}\,\delta_{\mu\nu}  \cr & +
{c_v\over 32\pi^2} \left[1+{\pi^2 \over 2}\right]\,\delta^{ab} \left (\vert q \vert \delta_{\mu\nu}+
{q_{\mu}q_{\nu} \over \vert q \vert} \right ) +
\cr & + \CO(\Omega^{-1},L^{-1},\Lambda^{-1}).\cr}}
In a similar way the contributions of  bosonic Pauli-Villars ghosts
$\phi_i$ read
\eqn\pvi{\eqalign {^{^{(PV)}}_{\phantom{a}(1)}\Gamma_{\mu \nu}^{ab}(q
)  = & {c_v\over 3\pi^2}\, \Omega\, m_i^2\,\delta^{ab}\,\delta_{\mu\nu}
\cr & -
{ c_v\over 4\pi}\,\Lambda \, m_i^2\,
\lambda_i^{-{1\over 2}}\, \delta^{ab}\,\delta_{\mu\nu}  \cr &-  {c_v\over
16\pi^2}\, m_i^2\, \delta^{ab}\, \left (\vert q \vert  \delta_{\mu\nu}  +
{q_{\mu}q_{\nu} \over \vert q \vert} \right )  \cr & +
\CO(\Omega^{-1},L^{-1},\Lambda^{-1}) ,\cr}
}
and
\eqn\pvii{\eqalign{ ^{^{(PV)}}_{\phantom{a}(2)}\Gamma_{\mu \nu}^{ab}(q) = & -
{c_v\over 6\pi^2}m_i(2m_i+1)\,\Omega \,\delta^{ab}\,\delta_{\mu\nu}
\cr & +
{c_v\over 4\pi}\,\Lambda\,  m_i^2\,
\lambda_i^{-{1\over 2}} \, \delta^{ab}\, \delta_{\mu\nu} \cr &
 + O(\Omega^{-1},L^{-1},\Lambda^{-1}) .\cr }
}
The Pauli-Villars fermionic ghosts $\psi_j$ and $\bar{\psi_j}$ give a similar
contribution with
an additional factor $(-2)$.

Therefore the cancellation of linear divergences requires that
\eqn\lin{2\sum_{j=1}^{M_f} n_j-\sum_{i=1}^{M_b} m_i
- 1 - (4n+1)  = 0.
}
The finite terms generated by the auxiliary pre-cutoff regularization
are not transverse as required by Slavnov-Taylor identities.
 Therefore, the
preservation those identities imposes more constraints on
the exponents of the Pauli-Villars regulators,
\eqn\gi{4\sum_{j=1}^{M_f} n_j^2 -2\sum_{i=1}^{M_b} m_i^2 + 1 -
(4n+3)^2 = 0 .}
It is easy to see that once the  condition \lin\ is
satisfied, there are not further divergences  provided that
$2n-n_j+1>0$ and $2n-m_i+1>0$ for every $i=1,\cdots M_b$ and
$j=1,\cdots M_f$. These conditions guarantee the absence of
subdivergences associated to the two-point functions of the
different ghost fields because the ultraviolet behavior of the
gluon propagator compensates the singularities associated to
 ghost-gluon interactions. On the other hand there are not divergences
 associated to three-point functions. In this case the potentially
logarithmic divergent terms cancel by algebraic reasons and the final
contribution remains finite.
 In the same way, it is easy to verify that
once the condition \gi\ is satisfied all the other Slavnov-Taylor
 identities of
the theory  are also satisfied. The reason is that only linear
divergences generate anomalous contribution to the Green's
functions. Anomalous contributions arise only in the
Slavnov-Taylor identities which state the transversality of the two
gluon function and was discussed above, and in the identity which
relates the  two-point and three-point gluonic
functions, which also vanishes if the condition \gi\  is satisfied \sing.
Of course there always exist an infinite number of solutions
 of the constraints \lin\gi\ but in order to keep
only  local interactions  all the exponents of the
Pauli-Villars interacting terms should  be integer numbers.
There is however a large number of solutions satisfying all these
requirements. The simplest one
$ m_1  =2; \, m_i=1, i=2,\cdots  ,8n^2 +4n+1;\, n_j=2,  j=1,\cdots ,2n^2 + 2n+1
$, is parametrized by an arbitrary
positive integer $n>0$.

	It is obvious from the above calculations that
the one-loop effective
action does not contains any pseudoscalar term and therefore
in this regularization scheme there are not perturbative corrections
to the wave
function normalization or the Chern-Simons coupling constant.
Therefore, we have found a gauge invariant regularization which
does not yield any renormalization effect on the coupling constant
 at one loop
level.  In particular, this implies that the symmetry arguments do not
enforce universality in the renormalization of $k$.
This behavior  is similar to the one obtained in Ref.
\ref\cero{E.Guadagnini, M. Martellini,  M. Mintchev,  Phys.Lett. { B 227}
(1989) 111}
 by means
of  a regularization which is not gauge invariant and presumably  the
non-renormalization of $k$ is also  preserved  up to two loops in the
present scheme.

 We  remark that although the above
 regularization method  is  correct from a perturbative viewpoint
(the correlations functions are finite and satisfy  Slavnov-Taylor
 identities),
 it might not have  a non-perturbative
meaning because  the regularized action is imaginary and
therefore the functional measure is not damped enough for large field
configurations. For such a reason, we shall restrict ourselves to
a pure perturbative analysis.

Once  we have shown that there is not universality in the
renormalization of $k$, we proceed to generalize the
regularization scheme to obtain different renormalizations of $k$.
This generalization can be implemented  by analogy with the
observation made by  L\"uscher and Coste about the
existence of a similar ambiguity in the effective action of $2+1$
massless fermions in the presence of a background gauge field
\lc. They showed that the  coefficient of the parity anomalous
Chern-Simons term of the effective action is determined modulo
a shift by $2\pi s$, associated to  the number of  ghosts
fields $s$ involved in a Pauli-Villars regularization or a parameter
in   different lattice
regularizations. In our case this ambiguity can be obtained
by means of  new Pauli-Villars ghost fields  interacting with
gauge fields by  pseudoscalar couplings.
The regularized action in this case would be given by
\eqn\final{\eqalign{S'_{reg}(A,\bar{c},
 {c},\phi, \bar {\psi}, {\psi}, \xi, \bar{\chi},
\chi)= & S_{reg}(A,\bar{c},
 {c},\phi, \bar {\psi}, {\psi})+{1\over 2}\sum_{r=1}^{N_b}
({\xi}_r, [I+ i{l_r\over \Lambda}\ast d^{\ }_A] \xi_r) \cr &+
\sum_{s=1}^{N_f} (\bar{\chi}_s, [I+ i{\eta_s\over \Lambda}\ast d^{\
}_A] \chi_s) }}
where the new ghost fields $\xi$ are bosonic, whereas $\bar{\chi}$
and
$\chi$ are grassmannian.

The contribution of the new bosonic ghost fields $\xi$ to the
two point
function
\eqn\axial{\eqalign {^{^{(PV)}}_{}{\Gamma'}_{\mu \nu}^{ab}(q) & =
-  {c_v\over 6\pi^2}\, \Omega\,\delta^{ab}\,\delta_{\mu\nu}   -
{c_v l^2_i\over 12\pi^2 \Lambda^2}\, \Omega\,
\delta^{ab} ( q^2 \delta_{\mu\nu} - q_{\mu}q_{\nu} )\cr
 & +
{c_v l_i\over 4\pi^2 \Lambda}\, \Omega\,\delta^{ab}\,
\epsilon_{\mu\nu\rho} q_\rho +
{c_v\over 16\pi^2}\delta^{ab} \left (\vert q \vert \delta_{\mu\nu} -
{q_{\mu}q_{\nu} \over \vert q \vert} \right )    \cr & -
{c_v l_r\over 4\pi \vert l_r \vert}\, \delta^{ab}
\epsilon_{\mu\nu\rho} q_\rho   +
\CO(\Omega^{-1},L^{-1},\Lambda^{-1}), \cr}  }
presents two new types of   divergent terms.

Further divergent contributions also arise in the three-point
and four-point functions,
\eqn\tres{\eqalign{\Gamma^{abc}_{\mu\nu\sigma}(q,r) & = -
{ic_v l_i\over 12\pi^2\Lambda}\, \Omega \, f^{abc}
\epsilon_{\mu\nu\sigma}   +
{ic_vl_i\over 12\pi\vert l_i \vert} f^{abc}
\epsilon_{\mu\nu\sigma}  \cr & -
{ic_v l^2_i\over 36\pi^2 \Lambda^2}\, \Omega\, f^{abc}
( (2q_{\nu} +
r_{\nu}) \delta_{\mu\sigma} - (q_{\mu}+2r_{\mu}) \delta_{\nu\sigma}
-  (q_{\sigma}-r_{\sigma}) \delta_{\mu\nu} )
  \cr  & +\CO(\Omega^{-1},L^{-1},\Lambda^{-1}),\cr}}
\eqn\cuatro{\eqalign{\Gamma^{abcd}_{\mu\nu\gamma\sigma}(q,r,s)&=
{c_v l^2_i \over 144\pi^2 \Lambda^2}{}\, \Omega \,
\lbrack \delta^{ab}\delta^{cd} (
\delta_{\mu\gamma}\delta_{\nu\sigma}
 + \delta_{\mu\sigma}\delta_{\nu\gamma}
-2\delta_{\mu\nu}\delta_{\sigma\gamma} ) +
 \delta^{ad}\delta^{bc} ( \delta_{\mu\gamma}\delta_{\nu\sigma}
 \cr &
 + \delta_{\mu\nu}\delta_{\sigma\gamma}
-2\delta_{\mu\sigma}\delta_{\nu\gamma} ) +
 \delta^{ac}\delta^{bd} ( \delta_{\mu\sigma}\delta_{\nu\gamma}
 + \delta_{\mu\nu}\delta_{\sigma\gamma}
-2\delta_{\mu\gamma}\delta_{\sigma\nu} ) \rbrack
\cr  & +\CO(\Omega^{-1},L^{-1},\Lambda^{-1}),\cr}}
because of the very singular ultraviolet behavior of the
longitudinal part of  propagators of the new Pauli-Villars fields.
Similar divergent contributions are generated by the new
fermionic ghost fields $\chi$. In fact, the contribution of those
fields only differ by a factor $(-2)$ of the  contributions \axial\ -
\cuatro\ of the $\xi$ fields. Notice,  however, that there are not
further divergences in the contribution of higher point functions.

Thus, if besides the normal condition for the
cancellation of  linear divergences \lin\  we impose the additional
conditions
\eqn\cancc{2{N_f} = N_b\qquad 2\sum_{s=1}^{N_f} \eta_s-\sum_{r=1}^{N_b} l_r  =
0 \qquad 2\sum_{s=1}^{N_f} \eta^2_s-\sum_{r=1}^{N_b} l^2_r  =
0}
on the  new Pauli-Villars
regulators all new types of divergences cancel out.
The dependence of the  Pauli-Villars conditions \cancc\
on the inverse  of the masses, $\l_i/\Lambda$ and $\eta_i/\Lambda$,
of the regulating
fields is due to the the fact that the leading ultraviolet behavior
of the new propagators is governed  by the
longitudinal modes, which behave
 as $\CO(1)$ in the large $q$-momentum  limit.

On the other hand, there is no change on the condition \gi\
which preserves Slavnov-Taylor identities because the new regulating
fields are Pauli-Villars fields and once the finiteness conditions
\cancc\ are satisfied   their contribution
is gauge invariant. Since there are not further constraints on
the values of the parameters $\l_i$ and $\eta_i$ any of the
infinite real solutions of  equations \cancc\
can be used to complete the definition of the regularization.

Collecting all the new finite contributions to the
pseudoscalar part of the effective action we get
\eqn\eff{{c_v\over k} \left\lbrace \sum_{r=1}^{N_b}
{l_r\over |l_r|}
  -2\sum_{s=1}^{N_f}{\eta_r\over |\eta_r|}
 \right\rbrace S_0(A).}
This implies that the renormalized coupling constant $k$ is shifted
by  an even multiple
of the dual Coxeter number $c_v$,
\eqn\shift{k+2  s c_v,}
where $2s$  is the even integer defined by the difference of
sums of $\pm 1$ in \eff.
A renormalization
with a shift by  an odd multiple of $c_v$ can be induced by
means of a slight
(geometric) generalization of the above regularization.
In any case, this result shows that in Chern-Simons theory the
degree of ambiguity in the renormalization of the coupling
constant is at least as large as in the case of the fermionic parity
anomaly. On the other hand, we remark the absence of
renormalization of the wave function, which is in contrast with the
results  obtained by other methods involving Yang-Mills terms
with higher covariant derivatives \pl\car.

In spite of the existence of the ambiguity which breaks the universality
properties
of the renormalization, it is quite remarkable that the integer character
of the coupling constant is always preserved by the above quantum corrections.
This fact  is not a consequence of the Slavnov-Taylor identities, which  are related to invariance under infinitesimal gauge transformations.
The analysis of this phenomenon deserves further investigation (see \new).

{\bf Acknowledgements:} {\it J. L. L. was supported by a MEC fellowship  (FPI
program)
and G.L. by a CONAI (DGA) fellowship. We also acknowledge to  CICYT
for partial financial support under grant AEN93-219}.\bigskip
\vfill\eject
\def\cistrefs{\vfill\immediate\closeout\rfile
\baselineskip=14pt\centerline{{\bf References}}
\bigskip{\frenchspacing
\escapechar=` \input \jobname.refs \vfill}\nonfrenchspacing}
%

\cistrefs
\end
\figures
\fig {1.} { One loop Feynman diagrams contributing to the 2-point function
involving gluon loops.}

\end
\vskip.2cm
{\hskip1,4cm \epsfxsize=10cm \epsfbox{figurafinal.ps}}
\centerline{{\bf Figure 1.}{ One loop Feynman diagrams contributing to the 2-point function
involving gluon loops.}
\vskip 4cm